\documentstyle[prl,aps,epsf,twocolumn]{revtex}
\begin{document}
\title{
CTRW Pathways to the Fractional Diffusion Equation.
}
\author{Eli Barkai\\
Department of Chemistry,\\
Massachusetts Institute of Technology,\\
Cambridge, MA 02139.\\
}
\date{\today}
\maketitle

\begin{abstract}

{\bf Abstract} 

The foundations of the  
fractional diffusion equation are investigated
based on coupled and decoupled 
continuous time random walks (CTRW).
For this aim we find an exact
solution of the decoupled CTRW,
in terms of an infinite sum of stable probability densities. 
This exact  solution is then used to understand the meaning and domain 
of validity of the fractional diffusion equation.
An interesting behavior is discussed for coupled memories  (i.e., 
L\'evy walks). The moments of the
random walk exhibit strong anomalous diffusion,
indicating (in a naive way) the breakdown of  
simple scaling behavior and hence of the fractional 
approximation. 
Still the Green function
$P(x,t)$ is described well by the fractional diffusion equation,
in the long time limit.

\end{abstract}

\section{Introduction}

 Fractional calculus is an old field of mathematical analysis
which deals with integrals and derivatives of arbitrary
order \cite{Samko,Hilfer1,Main1}. Fractional diffusion equations
were introduced to describe anomalous non Gaussian transport systems
\cite{Bochner,schneider,saichev,MBK,BS,Chechkin,Tran,Lutz,WZ} 
(see \cite{review} for review).
The stochastic foundation \cite{Hilfer,compte,soko,barkai9}
of these equations is the continuous
time random walk (CTRW) introduced by Montroll and Weiss \cite{Weiss1}.
The relation between CTRW and the fractional equations is the
reason for a renewed interest in the properties of CTRWs.

In this work we investigate the limitations and the
domain of validity of the fractional diffusion equation
based on coupled and decoupled CTRWs.
Some limitations on the fractional
framework were partially addressed in \cite{Tran,barkai9,Soko1},
for the sub diffusive case
in dimension $d$.
Here we consider the one dimensional
fractional diffusion equation \cite{saichev} 
\begin{equation}
{\partial^{\alpha} \over \partial t^{\alpha} } P_{fr}(x,t) = {1 \over 2}  
{\partial^\mu \over \partial |x|^\mu} P_{fr}(x,t)+{P_{fr}(x,0)t^{-\alpha}\over
\Gamma(1 - \alpha)},
\label{eqIn00}
\end{equation}
where ${\partial^{\alpha} \over \partial t^{\alpha} }$ is the 
fractional Riemann--Lioville (time) derivative of order $\alpha$
and ${\partial^\mu \over \partial |x|^\mu}$ is
the Riesz space fractional derivative of order $\mu$.
These fractional derivatives are integro-differential operators,
whose definition is given in \cite{Samko,saichev,review}.
The last term on the right hand side of Eq.  (\ref{eqIn00}) is the source term
which depends on initial conditions.
We consider free boundary conditions and initial conditions
concentrated on the origin $P_{fr}(x,0) = \delta(x)$, then the Fourier--Laplace
$(k$--$u)$ transform of the Green function is
\begin{equation}
P_{fr}(k,u) = { u^{\alpha - 1} \over u^{\alpha} + |k|^\mu /2 }.
\label{eqIn01}
\end{equation}
This equation is for our purposes the definition of the fractional
equation (\ref{eqIn00}).   The inversion of Eq. (\ref{eqIn01})
yields
\begin{equation}
P_{fr}(x,t)=t^{-\alpha/\mu} K\left({ x \over t^{\alpha/\mu} } \right),
\label{eqIN000}
\end{equation}
where $K(z)$ is a scaling function whose properties are 
given in \cite{saichev,Main}.
The probability interpretation of $P_{fr}(x,t)$ is restricted
to $\{ 0 <  \mu \le 2 \}\cap  \{0 < \alpha \le 1\}$ and
$\{ 1 < \alpha\le \mu \le 2\}$ \cite{Main}.
When $\mu=2$ and $\alpha=1$ the fractional equation reduces
to the ordinary Gaussian diffusion equation, 
if $\mu<2$ and $\alpha=1$ it describes L\'evy
flights. 
When $\mu=2$ the equation 
describes sub or enhanced diffusions, $\langle x^2 \rangle \propto t^{\alpha}$,
according to $\alpha<1$ or $\alpha >1$, respectively.

Eq. (\ref{eqIn01}) has a long history:
for certain values of $\alpha$ and $\mu$ it 
was derived
from the CTRW model
\cite{saichev,Weiss1,Tunaley1,Wong,KBS,Blumen},
using the long wave length $k \to 0$ small $u \to 0$
approximation (see details below). 
This approximation
was used many times to investigate the long time behavior
of the CTRW. It is based on
the simplifying assumption 
that the scaling behavior Eq. (\ref{eqIN000}) 
holds
\cite{Tunaley1}.
In \cite{Marcin,Marcin1}
a rigorous approach, based on limit theorems, was used
 to classify
the asymptotic behaviors of different types of
CTRWs. The work in \cite{Marcin,Marcin1} is in agreement with Eq. 
(\ref{eqIn01}) and the older work in this field
(i.e., again for certain $\alpha$ and $\mu$ and see details below). 

 Here exact solution
of the decoupled CTRW in $(x,t)$ space is found
in terms of an infinite sum of stable probability densities.
This exact solution is used 
to investigate the meaning and limitations
of the fractional diffusion equation.
For example:  we show that certain
solutions of the fractional diffusion equation diverge on the
origin, a behavior not found in the corresponding
CTRW.
We also show that certain CTRW solutions, converge extremely slowly
toward the fractional diffusion approximation.

The exact solution of the CTRW is based on a particular 
choice of waiting time and jump length distributions. Beginning
in Sec. \ref{SecG} we investigate the fractional
approximation using a more general approach.
Both coupled and decoupled CTRWs are considered.
As far as I know, the relation between coupled CTRWs and
the fractional diffusion equation was not discussed previously.
In Sec. (\ref{secSub}) we discuss 
Castiglione et al's \cite{Vulp} 
objection to the fractional
diffusion equation, for systems exhibiting
strong anomalous diffusion.

\section{CTRW an Exact Solution}
\label{SecEXACT}

 In this  section we find an exact solution of the decoupled, one
dimensional, CTRW
model in terms of an infinite sum of stable functions.
Usually the solution of the CTRW, for finite times,
is found using a numerical approach. 
The exact solution is used to understand the meaning and
limitations
of the fractional diffusion equation.

 For the well known decoupled  CTRW model,
a particle is trapped on the origin for time $t_1$, it
is then displaced  to $x_1=\delta x_1$ then the particle
is trapped for time $t_2$ and then it jumps again,
the process is  then renewed.
Let $\psi(t)$ be the probability density function (PDF) of the independent
identically distributed (IID)
random variables $\{t_i\}$ $i=1,2,\cdots$, while the IID
displacements $\{ \delta x_i \}$ are described by a PDF
$f(\delta x)$.  The displacement $\delta x_i$ is related to the
coordinate of the particle according to $\delta x_i= x_i - x_{i-1}$
and $x_0=0$. Here it is assumed
that start of observation is also
start of the process. 

 We assume that the 
PDF of waiting times $\psi(t)=l_{\alpha,1}(t)$
is a one sided stable probability
density. Namely, 
its Laplace $t \to u$ transform is
$\psi(u) = \exp( - u^{\alpha})$,
and $0<\alpha <1$. The PDF of jump
lengths is chosen to be a symmetric
stable density $f(\delta x)= l_{\mu, 0}(2^{1/\mu} \delta x)$, namely
its Fourier transform is $f(k)=\exp( - |k|^{\mu}/2)$ 
and $0<\mu \le 2$. The case $\mu=2$ (i.e.,
Gaussian jumps), $0<\alpha <1$ and  dimension $d\ge 1$
was considered in \cite{Tran}. Properties of the stable densities
can be found in \cite{Sch1,Feller,Bendler,remark2}. In the following sections
the more general case where $f(\delta x)$ 
and $\psi(t)$ belong to the domain of attraction of the L\'evy stable
laws, as well as coupled space time memories, is considered.

\begin{figure}[htb]
\epsfxsize=20\baselineskip
\centerline{\vbox{
      \epsffile{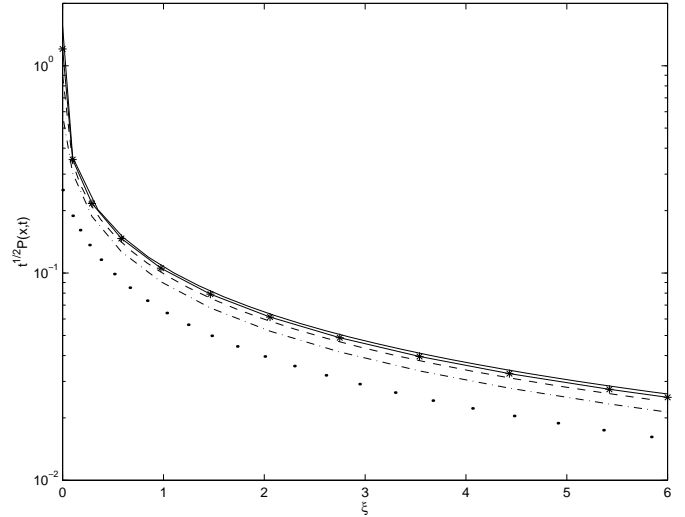}  }}
\caption {
We show $t^{1/2} P(x,t)$ versus the scaling
variable $\xi=x^2/t$ for the CTRW process.
The curves in the figure are for the times
$t=1,5,25,125$ (dots, dot dash, dashed, stared curve) respectively.
We see that the CTRW solution converges in the limit of large times
to the 
fractional diffusion approximation (solid line). 
The fractional solution diverges on the origin, hence
it is cutoff in the figure. 
To find the exact solution we
used Cauchy $\mu=1$ and Smirnov $\alpha=1/2$ stable laws.
The results shown in Figs. $1-4$ were obtained using Mathematica.
}
\label{fig2}
\end{figure}

 Because the model is decoupled \cite{Weiss1}
\begin{equation}
P(x,t) = \sum_{s=0}^{\infty} N_{CT}(s,t) W(x,s),
\label{eqEx01}
\end{equation}
where $N_{CT}(s,t)$ is the probability
that $s$ steps are made in time
interval $(0,t)$ and $W(x,s)$ is the PDF that
the random walk is on $x$  after $s$ steps. 
Using the convolution property of the symmetric stable densities,
it is easy to show that 
\begin{equation}
W(x,s) = \left({ 2^{1/\mu} \over s^{1/\mu}} \right)
 l_{\mu, 0}\left({2^{1/\mu} x\over s^{1/\mu}} \right).
\label{eqEx02}
\end{equation}
$N_{CT}(s,t)$ is found using the convolution
theorem of Laplace transform (see details in \cite{Tran})
$$ N_{CT}(0,t) = 1 - L_{\alpha,1}(t),$$
\begin{equation}
N_{CT}(s,t) =L_{\alpha,1}\left({ t \over s^{1/\alpha} } \right) 
-L_{\alpha,1}\left({ t \over (s+1)^{1/\alpha} } \right) 
\label{eqEx03}
\end{equation}
and $L_{\alpha,1}(t)\equiv \int_0^t l_{\alpha,1}(t) dt$ is the one sided 
stable distribution function.
Hence from Eqs. (\ref{eqEx01}-\ref{eqEx03})
$$ P(x,t) = \left[ 1- L_{\alpha,1}\left( t \right) \right] \delta(x) + $$
$$\sum_{s=1}^{\infty} \left[ L_{\alpha, 1}\left({ t \over s^{1/\alpha} }\right)-
L_{\alpha, 1}\left({ t \over (s+1)^{1/\alpha} } \right)\right] \times $$ 
\begin{equation}
\left({ 2^{1/\mu} \over s^{1/\mu}} \right)
 l_{\mu, 0}\left({2^{1/\mu} x\over s^{1/\mu}} \right).
\label{eqEx04}
\end{equation}
The first term on the right hand side describes
random walks for which the particle did not
leave the origin within the observation time $t$;
the other terms describe random walks where the number of steps is $s$.
In Fig. (\ref{fig2}) we show an exact solution of the CTRW process,
in a scaling form, for the case $\alpha=1/2$ and $\mu=1$. 

 The fractional diffusion approximation is reached when the summation
in Eq. (\ref{eqEx04}) is replaced with integration. We show
below that such a replacement is not always valid.
Using 
$$ L_{\alpha,1}\left( {t \over s^{1/\alpha} } \right) -
 L_{\alpha,1}\left( {t \over (s +1)^{1/\alpha}} \right) \simeq - {\partial \over \partial s}  L_{\alpha,1}\left( {t \over s^{1/\alpha} } \right) ds = $$
\begin{equation}
{1 \over \alpha} {t \over s^{1/\alpha+1}} l_{\alpha, 1}\left( {t \over s^{1/\alpha}} \right),
\label{eqEx05}
\end{equation}
and neglecting the delta function contribution in Eq. 
(\ref{eqEx04}) we find
\begin{equation}
P(x,t)  \simeq \left( { 2^{{1\over \mu} } t \over \alpha } \right) \int_0^{\infty} d s s^{ -{1 \over \mu} - {1 \over \alpha}-1}  l_{\alpha, 1}\left( {t \over s^{{1\over \alpha}}}
 \right)
 l_{\mu, 0}\left({2^{{1\over \mu}} x\over s^{{1\over \mu}}} \right),
\label{eqEx05a}
\end{equation}
this approximation might be expected to work well only in the long
time limit.
The right hand side of  Eq. (\ref{eqEx05a}) is the integral
 solution of the fractional
diffusion equation
(\ref{eqIn00}) obtained by Saichev and Zaslavsky \cite{saichev}
[i.e., only for $0<\alpha \le 1$ and $0<\mu\le 2$ and see Eq.
(\ref{eqAp06}) in Appendix A].
In subsection \ref{subsub} we will show that in the vicinity of the origin $x=0$
and  for $\mu\le 1$ the fractional approximation Eq. (\ref{eqEx05a})
does not work well. Let us therefore analyze the sum Eq. (\ref{eqEx04}) more carefully.

First we rewrite Eq. (\ref{eqEx04}) 
\begin{equation}
P(x,t) = \sum_{s=0}^{\infty} W(x,s) \int_{t/(s+1)^{1/\alpha}}^{t/s^{1/\alpha}} l_{\alpha,1}\left( t' \right) d t' 
\label{eqEx01a}
\end{equation}
where $W(x,0)=\delta(x)$. Using the Euler-Mclaurin summation
formula \cite{Abra} we have
$$ P(x,t) = \delta(x) \int_t^\infty l_{\alpha,1}(t') d t' + {1 \over 2} W(x,1)\int_{t/2^{1/\alpha}}^t l_{\alpha,1}(t') d t' + $$
\begin{equation}
\int_1^{\infty} W(x,s) \left[ \int_{t/(s+1)^{1/\alpha}}^{t/s^{1/\alpha}} l_{\alpha,1}\left( t' \right) d t'\right] d s + \cdots,
\label{eqEx02a}
\end{equation}
%
where $\cdots$ are the higher order terms in the Euler--Mclaurin
formula.
As discussed below, important correction terms to the fractional
diffusion approximation 
can be calculated based on Eq.
(\ref{eqEx02a}).
 In the long time limit the first two terms on the right hand
side of Eq. (\ref{eqEx02a}) decay like $t^{-\alpha}$. Within the
fractional diffusion approximation these terms are neglected.
The third term on the right hand side of Eq. (\ref{eqEx02a}) 
yields the leading contribution to $P(x,t)$ in the long time limit. 
For this term only contributions from large $s$ are
important, when $t \to \infty$. Using
\begin{equation}
\int_{t/(s+1)^{1/\alpha}}^{t/s^{1/\alpha}} l_{\alpha,1}\left( t' \right) d t' \sim l_{\alpha,1}\left({t \over s^{1/\alpha}}\right){t \over \alpha s^{1/\alpha + 1}}
\label{eqEx03a}
\end{equation}
we find in the limit $t \to \infty$
\begin{equation}
P(x,t) \sim \int_1^{\infty} W(x,s) {t \over \alpha s^{1/\alpha +1 }} l_{\alpha,1}\left( { t \over s^{1/\alpha}} \right) ds.
\label{eqEx04a}
\end{equation}
This expression differs from the exact solution of the fractional diffusion
equation by its nonzero lower limit in the integral.  Comparing Eqs. 
(\ref{eqEx05a}) and (\ref{eqEx04a}) we see that the shortcoming of the
fractional approximation is that it attempts to give statistical weight
to trajectories where number of jumps
is ``less than one''. Subtracting Eq. (\ref{eqEx04a}) from Eq. (\ref{eqEx05a}),
and using $l_{\alpha,1}(t) \propto t^{-(\alpha + 1)}$ when $t\to \infty$ we
have
$$
\int_0^1 W(x,s) {t \over \alpha s^{1/\alpha + 1}} l_{\alpha,1} \left({ t \over s^{1/\alpha}} \right) ds \propto t^{-\alpha} \int_0^1 W(x,s) d s. 
$$
The integral $\int_0^1 W(x,s) ds$ may become very large when $x$ is small,
and when $x=0$ the integral may diverge (i.e., since $W(x,s)|_{x=0}\propto s^{-1/\mu}$) . Then the convergence of the CTRW to
the fractional approximation becomes extremely slow and when $x=0$
the fractional approximation breaks down for $\mu \le 1$.

%

\subsection{Diverging Solution of the Fractional Equation}
\label{subsub}

 We now investigate in detail the behavior of the CTRW and the corresponding
fractional diffusion equation at the origin $x=0$. 
In Appendix A,  the solution of the fractional
diffusion equation
is used to show that
\begin{equation}
P_{fr}(x,t)|_{x=0}=   
\left\{
\begin{array}{c c}
2^{1 \over \mu} t^{-\alpha/\mu} 
{\Gamma\left(1 - 1/\mu\right)\over \Gamma\left(1-\alpha/\mu\right) } l_{\mu,0}(x)|_{x=0}  \ 
& \mu>1 \\
\ & \ \\
\infty & \mu\le 1, 
\end{array}
\right.
\label{EqL06}
\end{equation}
where $l_{\mu, 0}(x)|_{x=0}=\Gamma(1/\mu)/(\mu \pi)$. The
subscript fr stands for fractional diffusion approximation.
 We assumed $\alpha<1$ since
the case $\alpha=1$ yields stable propagator $P_{fr}(x,t)$ which
does not diverge on the origin (see Appendix A).

\begin{figure}[htb]
\epsfxsize=20\baselineskip
\centerline{\vbox{
      \epsffile{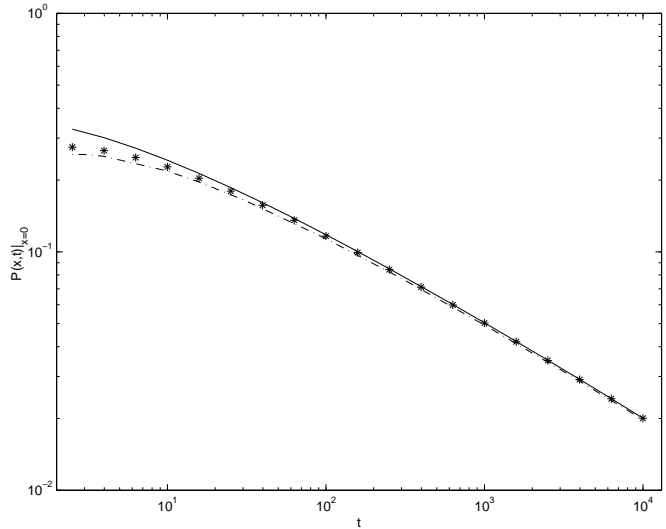}  }}
\caption {
The behavior of exact CTRW on the origin,
$P(x,t)|_{x=0}$ versus $t$ on log log plot
for the case $\alpha=1/2$ and $\mu=1$.
Not shown is the delta function contribution. The stars are the
exact solution  (\protect{\ref{eqEx04}})
while the solid curve is the asymptotic behavior Eq. 
(\protect{\ref{eqEx06}}). The fractional diffusion equation yields
for this case $P(x,t)|_{x=0}= \infty$ and hence is invalid. 
The approximation based on the Euler--Mclaurin formula,
 Eq. (\protect{\ref{eqEx02a}}), is the dot dashed curve. 
It yields good agreement with the exact results. 
}
\label{fig1}
\end{figure}

 In Appendix B the exact CTRW solution,  Eq.
(\ref{eqEx04}), is used to find the behavior of the CTRW on the origin
(i.e. for the non singular terms).
For $\alpha<1$ we find
\begin{equation}
P(x,t)|_{x=0} \sim
 \left\{
\begin{array}{c c}

2^{1/\mu} 
{\zeta\left({1 \over \mu} \right) \over \Gamma(1 - \alpha) } {1 \over t^{\alpha} } l_{\mu,0}(x)|_{x=0}  & \mu< 1 \\
\ & \ \\
{2 \alpha \over \pi \Gamma(1 - \alpha) } {\ln(t) - \hat{\psi}(\alpha) \over t^{\alpha} } & \mu=1 \\
\ & \ \\
2^{1/\mu} 
 {\Gamma\left( 1 - 1/\mu\right)\over\Gamma\left(1-\alpha/\mu\right) } 
{1 \over t^{\alpha/\mu }} l_{\mu,0}(x)|_{x=0}  & \mu> 1, 
\end{array}
\right.
\label{eqEx06}
\end{equation}
where $\zeta(z)$ is the Riemann zeta function and 
$\hat{\psi}(\alpha)$ is the psi function [of course not related
to $\psi(t)$].
In Fig. (\ref{fig1})  we show $P(x,t)|_{x=0}$ versus
$t$ for $\mu=1$, exhibiting how  the exact CTRW solution converges to
its asymptotic limit at the origin. 
Comparing Eq. (\ref{eqEx06}) with Eq. (\ref{EqL06}) we see 
that the infinity found for $\mu\le 1$ within the fractional
framework  is not related to
the underlying CTRW. As mentioned, this shortcoming within the
fractional approximation is due to the fact that
number of steps  in the random walk is an integer
which cannot generally be approximated with a continuum approach
[i.e., replacement of summation with integration in Eq. 
(\ref{eqEx04})
is not justified for $0<\mu\le 1$ in the vicinity of the origin].
In Fig. \ref{fig1} we also show the approximation based on the
Euler-Mclaurin formula Eq. (\ref{eqEx02a}). In contrast to
the fractional diffusion approximation, Eq. (\ref{eqEx02a})
yields good agreement with the exact results.
Note that divergence of the solution of the
$d>1$ dimensional
fractional diffusion equation with $\mu=2,\alpha<1$ (i.e. sub diffusive
case) at the origin was discussed in \cite{Tran,HilferD}. For these
cases the exact CTRW solution is a valuable tool.

%

\subsection{Slow Convergence Toward Fractional Approximation}
\label{subsub1}

\begin{figure}[htb]
\epsfxsize=20\baselineskip
\centerline{\vbox{
      \epsffile{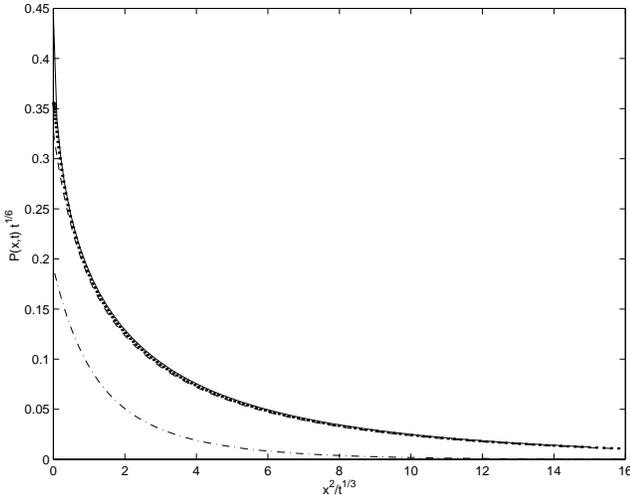}  }}
\caption {
The long time convergence of the exact CTRW solution
toward the fractional diffusion approximation for
$\alpha=1/3$ and $\mu=2$. We present $t^{1/6} P(x,t)$ versus the scaling
variable $x^2/t^{1/3}$.
The CTRW curves in the figure are for the times
$t=125,625,3125$ (dot dash, dashed, dot) respectively,
the solution of the fractional diffusion equation is the solid line. 
To obtain the exact CTRW solution we
use the waiting time density 
$l_{1/3,1}(t)=1/(3 \pi)t^{-3/2}K_{1/3}\left({2 \over \sqrt{ 27 t}} \right)$,
where $K_{1/3}$ is the modified Bessel function of the
second kind. The jump length probability density
is Gaussian $f(\delta x)= (4 \pi)^{-1/2}\exp(-\delta x^2/4)$. For $\alpha=1/3$ and $\mu=2$
the solution of 
the fractional diffusion equation is 
$P(x,t)=3|x|^{-1}\xi l_{1/6,1}(\xi)$ with $\xi=t/|x|^{6}$.
}
\label{fig3}
\end{figure}

\begin{figure}[htb]
\epsfxsize=20\baselineskip
\centerline{\vbox{
      \epsffile{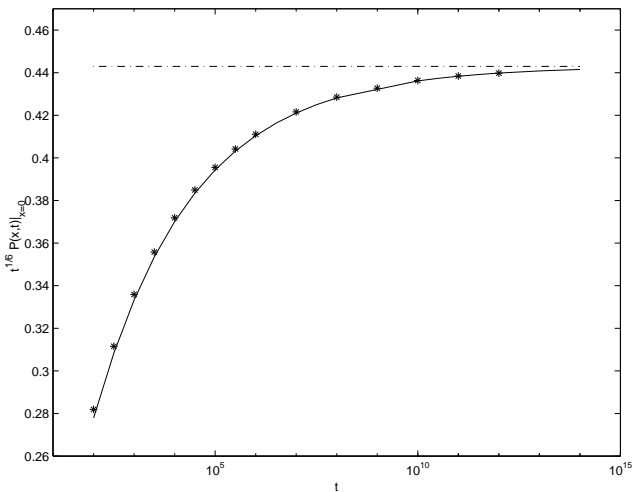}  }}
\caption {
The very slow convergence of the non singular CTRW solution toward
the fractional approximation on the origin. We use the same
parameters as in Fig.
\protect{\ref{fig3}}. The exact CTRW solution are the stars, the dot dash line
is the fractional approximation, and the solid curve is the approximation
based on
the Euler--Mclaurin formula,
Eq. (\protect{\ref{eqEx02a}}).  
}
\label{fig4}
\end{figure}

 Let us consider as an example the case $\alpha=1/3$ and $\mu=2$.
In Fig. \ref{fig3} we show the exact CTRW solution in scaling form.
As expected, for long times the CTRW solution seems
to converge toward the solution of the fractional diffusion
equation, though clear deviations of the CTRW solution
from the fractional approximation are seen in the vicinity of the origin.
 In Fig. \ref{fig4}, a closer look at the behavior at the origin 
is presented.  The figure shows that the CTRW convergence toward
the fractional diffusion approximation is extremely slow,
for $t=10^{13}$ deviations from asymptotic behavior are still
observed (note that since $\psi(u)=\exp(-u^{1/3})$
the natural time unit is $1$, though the mean time between jumps
diverges). Our improved approximation, Eq. 
(\ref{eqEx02a}), yields a good description of the
underlying CTRW for intermediate and long times. We note that as $\alpha \to 0$
convergence of the CTRW solution toward the fractional approximation
is expected to become much slower. And of course when $\alpha \to 1$, convergence becomes
faster, though then deviations from Gaussian behavior $(i.e.,
\mu=2,\alpha=1$) become
small. 

\section{The General Approach}
\label{SecG}

 While the exact solution presented in previous section
gives insight into the validity of the fractional diffusion
equation it is based on a particular choice of 
$\psi(t)$ and $f(\delta x)$. Here we shall consider a more general
approach.

 Let $P(x,t)$ describe a normalized  Green function of an unspecified
one dimensional random walk;
later we
treat specific examples in some detail. 
We shall use the convention that the arguments in
the parenthesis define the space we are working in,
thus $P(k,u)$ is the Fourier--Laplace transform of the Green
function $P(x,t)$. 
It is assumed that $P(k,u)$ is known exactly as is the case
for different types of CTRWs and for generalized master equations
of the type investigated in \cite{Metzler}.

Consider the expansion
\begin{equation}
P(k,u) = \sum_{n=0}^{\infty}  M_n(u) {(i k )^n\over n!},
\label{eq00}
\end{equation}
where it is assumed that all the moments of the random walk
\begin{equation}
M_n(u)=\left( {d \over i d  k}\right)^n P(k,u)|_{k=0}
\label{eq01}
\end{equation}
are finite. The case when the moments diverge will be discussed later.
According to Tauberian theorems \cite{Weiss1,Feller}, 
the small $u$ behavior of the moments 
in the Laplace domain yield the long time behavior of
the moments in the time domain. Namely, if
$M_n(u) \sim c_n u^{ - \xi_n -1}$ when $u \to 0$ then
$M_n(t) \sim c_n t^{\xi_n}/\Gamma( 1 + \xi_n)$ for $t \to \infty$,
where $\xi_n>0$ and $c_n$ are constants.
Our goal is to find the asymptotic moment generating
function $P_{AMG}(k,u)$ which contains all the information
on the $u \to 0$ behavior of the moments. 
We define this function according to
\begin{equation}
P_{AMG}(k,u) \equiv \sum_{n=0}^{\infty} c_n u^{ - \xi_n - 1} {\left( i k \right)^n \over n!}.
\label{eqMG}
\end{equation}
If this series can be summed (see examples to follow)
the function $P_{AMG}(k,u)$
yields in a compact way all the asymptotic 
information on the moments of the
underlying transport process.  If the inverse Laplace--Fourier
transform of $P_{AMG}(k,u)$ is a normalized non-negative
probability density, then it is safe to say that $P_{AMG}(k,u)$
is the Fourier-Laplace transform of the asymptotic
Green function i.e., $P(x,t)$ in the long time limit.
This is the case for most Gaussian transport systems.
Below we discuss  L\'evy walks where
$P_{AMG}(k,u)$ does not yield the Green function
$P(x,t)$ in the long time limit, 
even though it does contains all the information on
the long time behavior of the moments.

\section{Sub--Diffusion}
\label{SecS}

We now consider as an example the well known decoupled continuous time
random walk in the sub diffusive regime.
Similar to previous work (e.g. \cite{Weiss1})
we assume
\begin{equation}
\psi(u)=1 -  A u^{\alpha} + c_1 u^{2 \alpha} \cdots
\ \ \   0<\alpha<1,  
\label{Eq01S}
\end{equation}
for $u \to 0$, 
so that  $\psi(t) \propto t^{-(1 + \alpha)}$ for $t \to \infty$; hence
$\psi(t)$ is moment-less.
In what follows we use $A=1$ restoring $A$ only when it is important.
We also assume 
\begin{equation}
f(k) = 1 - { k^2 \over 2} +{ m_4 k^4 \over 4!} - {m_6 k^6 \over 6!} \cdots,
\label{Eq01Sa}
\end{equation}
where $m_i$ are the finite moments of the PDF $f(\delta x)$.
We assumed the $f(\delta x)$ is symmetric hence $m_{2 i +1}=0$,
the second moment being $m_2=1$. As discussed below some of these
assumptions can be relaxed.

 The Green function $P(x,t)$ of finding the random walker at $x$ at time $t$
is given in Fourier--Laplace space according to \cite{Weiss1}
\begin{equation}
P(k,u) = {1 - \psi(u) \over u} {1 \over 1 - \psi(u) f(k) }.
\label{Eq02S}
\end{equation}
The long time behavior of this equation is usually investigated
based on the long wave length approximation (e.g., \cite{review}),
namely by inserting
\begin{equation}
\psi(u) f(k) = 1 - u^{\alpha} - {k^2 \over 2}
\label{eq0034}
\end{equation}
and $\psi(u) =1 - u^{\alpha}$ in Eq. (\ref{Eq02S}). This approach
implicitly assumes that simple scaling Eq. 
(\ref{eqIN000}) with $\mu=2$ holds. Let us now see
why this is the case.

We expand $P(k,u)$ in $k$
$$ P(k,u) = {1 \over u} \left\{ 1 - \Omega(u) {k^2 \over 2} + 
 \left[6 \Omega^2(u) + m_4 \Omega(u)\right] {k^4 \over 4!} - 
\right. $$
\begin{equation}
\left. \left[  90 \Omega^3(u) + 30 m_4 \Omega^2(u) + m_6 \Omega(u) \right] {k^6 \over 6!} \cdots  \right\}
\label{Eq03S}
\end{equation}
where $\Omega(u) = \psi(u)/[1 - \psi(u)]$.
The $i$ th  term in the expansion gives the moment $M_i(u)$ of the random 
walker in-terms of the ``microscopic" moments $m_i$ and 
$\Omega(u)$. For example $M_0=1/u$ which means that the normalization is
conserved, $M_2(u)=\Omega(u)/u$ etc. In the long wave length approximation
(or in the fractional diffusion equation approach) one sets $m_i=0$ for
$i\ge 4$. To see why and when this works well we must consider the high
order moments $M_i$ and $i\ge 4$. For example
\begin{equation}
M_4(u)= {\left[ 6 \Omega^2(u) + m_4 \Omega(u) \right]\over u}.
\label{Eq03Sa}
\end{equation}
We now consider the $u\to 0$ limit of this expression since
this limit will yield the asymptotic expression for $M_4(t)$ 
when $t \to \infty$.
Using $\Omega(u) \sim 1/u^{\alpha}$ it is easy to see that
\begin{equation}
M_4(u)\sim {6 \over u^{1 + 2 \alpha}}, 
\label{Eq05S}
\end{equation}
and we see that $M_4(u)$ is independent of $m_4$ in the limit $u \to 0$. 
Similar behavior is found for all the higher order moments
\begin{equation}
M_{2 j}(u ) \sim {1 \over u} { \left( 2 j \right)! \over \left(2 u^{\alpha}\right)^{ j} }, 
\label{EqSO6}
\end{equation}
which is independent of $c_1$ and $m_{2 j}$ for $j>1$.
This behavior is
similar in some sense
to normal (i.e., $\alpha =1$) random walk 
 where all 
the moments $M_n(u)$ converge in a limit to simple Gaussian behavior 
which is independent
of the details of the underlying random walks. However now we
are not considering Gaussian diffusion.
It is also easy to show that behavior in Eq. (\ref{EqSO6}) is compatible
with the scaling assumption Eq. 
(\ref{eqIN000}).
Using Eq. 
(\ref{eqMG}) one finds 
\begin{equation}
P_{AMG}(k,u) = \sum_{j=0}^{\infty} { k^{2 j} (-1)^j \over u ( 2 u^{\alpha})^j},
\label{EqSO7}
\end{equation}
summing this geometric series we have
\begin{equation}
P_{AMG}(k,u) = { u^{\alpha - 1} \over u^{\alpha} + k^2 /2 }. 
\label{EqS08}
\end{equation}
Or we may inverse Laplace transform Eq. (\ref{EqSO7}) term by term
and find
\begin{equation}
P_{AMG}(k,t) = \sum_{j=0}^\infty \left( { - k^2 \over 2} \right)^j { t^{\alpha j} \over \Gamma\left( 1 + \alpha j \right)} = E_{\alpha}\left( - {k^2 t^{\alpha} \over 2} \right),
\label{eqS09a}
\end{equation}
where $E_{\alpha}(x)$ is the Mittag-Leffler function.
Since Eq. (\ref{EqS08}) is the Fourier Laplace transform
of a non negative probability density (see e.g. Appendix A),
it yields $P(x,t)$ in the long time limit.

Eq. (\ref{EqS08}) is the Fourier--Laplace transform of the 
fractional diffusion
equation 
(\ref{eqIn00}) when $\mu=2$ and $0<\alpha < 1$, 
and  well known 
within the CTRW community \cite{Tunaley1}.
The inverse Fourier--Laplace transform of Eq. (\ref{EqS08})
was investigated in \cite{schneider,Tran,Gurt} in dimensions $d=1,2,3$.
Here we showed that: $(i)$ this equation does indeed describe the
long time behavior
of the moments of the random walk to all orders 
and $(ii)$ that these moments depend only on three  parameters
of the model $\alpha,m_2=1$ and $A=1$ (i.e., universality) . 
Our approach clarifies the usual long wave length
approximation which is based on the exact
calculation of only the first two moments.
In the following section we will discuss
coupled memories where the asymptotic behavior 
is not as straightforward as for the decoupled case.

 In our derivation  we
assumed that start of observation and start of the process
coincide.   If the first step is described by
$\psi_1(t) \ne \psi(t)$,
one can show that our results are still valid
for $\psi_1(t)$ decaying faster then $\psi_1(t) \propto t^{-(1 + \beta)}$,
with $\beta>0$. When $\beta<< \alpha$ the convergence becomes slow.
If the random walk is biased, $m_1 \ne 0$, one can easily show
that biased fractional diffusion equation (e.g. \cite{Tran})
holds in the long time limit.

\section{Enhanced--Diffusion}
\label{SecE}

 We now consider an example exhibiting enhanced, {\em  L\'evy walk}
type of diffusion. We start by introducing the coupled
CTRW jump model, investigated by Zumofen, Klafter and Shlesinger
 \cite{ZK,KZS}
in the context
of chaotic maps. 
Such a random walk
is also related to transport in random
media \cite{LevLor,Levitz}, tracer diffusion in turbulent flow \cite{Swinney}
and to the blinking of Quantum dots \cite{Jung}.
Closely related models are the velocity models investigated in 
\cite{ZK,Mas,Araujo,BF}.

 In CTRW the random walk is entirely specified
by $\tilde{\psi}(\delta x, t)$, the probability density
to move a distance $\delta x$ in time $t$ in a
single jump event. For the jump model a coupled 
space-time memory is assumed
\begin{equation}
\tilde{\psi}\left(\delta x, t \right) = {1 \over 2} \delta\left( |\delta x| - t\right) \psi(t). 
\label{eqE01}
\end{equation}
Such a model describes a particle trapped on the origin
for time $t_1$
then it jumps to a new location whose distance from the 
origin is $|\delta x|=t_1$ (i.e, $x_1=\pm t_1$ with equal
probability), then the process is renewed.
From Eq. (\ref{eqE01}) we see that the random times
$\{ t_i \}$  are distributed according to
$\psi(t)$ and the length of each jump $i$
is $|\delta x_i| = t_i$.
Hence a large jump will ``cost" a long time. This is different
from the L\'evy flight model where jumps on all scales
are performed at constant time intervals.
Thus in some applications L\'evy walks are considered
more physical than  L\'evy flights; 
however, as discussed below these two models
are in fact deeply related.
The space time coupling in  Eq. (\ref{eqE01}) guarantees
that for the L\'evy walk model, $P(x,t)=0$ for $|x|>t$,
this in turn implies that all
moments of the random walk are finite.
For L\'evy flights even moments diverge.

 The Laplace--Fourier transform of the L\'evy walk
Green function is
\cite{ZK}
\begin{equation}
P(k,u) = {1- \psi(u) \over u \left[ 1-  \tilde{\psi}\left( u ,k\right)\right] }
\label{eqE02}
\end{equation}
where
\begin{equation}
\tilde{\psi}(k,u)=\int_0^{\infty} d t  e^{-ut} \cos\left( k t \right) \psi(t).
\label{eqE03}
\end{equation}
The moments of the random walk are now calculated using
Eq. (\ref{eqE02}) and Mathematica, one finds 
$ M_0(u) = 1/u $
$$ M_2(u) = {1 \over u} {\psi^{(2)}(u) \over 1 - \psi(u) } $$
\begin{equation}
 M_4(u) = {1 \over u} {\left[1 - \psi(u) \right] \psi^{(4)}(u) + 6 \left[ \psi^{(2)} \left( u \right) \right]^2 \over \left[1 - \psi(u)\right]^2 } 
\label{eqE07z}
\end{equation}
%
%
where $\psi^{(2 j)}(u)$ is the $2 j$ th derivative of $\psi(u)$ with 
respect to $u$.
Odd moments vanish due to the assumed symmetry of the random walk. Higher order
moments are calculated in a similar way, for the sake of space they are not
included here.

$ \  $

\subsection{Sub--Ballistic Enhanced Diffusion}
\label{secSub}

 Let as now consider $\psi(u) \sim 1 - u \tau + B u^{\beta} \cdots$ and
$1< \beta <2$. Unlike the previous example,
now the mean waiting time $\tau$ is finite and the second moment
of the waiting time distribution diverges. 
The model exhibits enhanced diffusion
$\langle x^2 \rangle \propto t^{\alpha}$ and $\alpha=3-\beta$.

 As we shall see in detail, the model exhibits a strong type of
anomalous diffusion. By definition \cite{Vulp} strong anomalous
diffusion behavior exhibits $M_{2j}(t) \propto t^{f(j)}$ where
$f(j)$ is a non--linear function of $j$ (see related work
\cite{LevLor,Andersen,Car}). 
Castiglione et al \cite{Vulp} point out that strong anomalous diffusion
implies the failure of the
standard scaling assumption 
Eq. (\ref{eqIN000}), since this equation predicts
$M_{2j}(t) \propto t^{\alpha j}$ a behavior
called weak anomalous diffusion.

Castiglione et al argue quite generally that a 
dynamical system exhibiting strong anomalous diffusion
cannot be described by
fractional diffusion equation. 
We show below, based on work of Zumofen et al
\cite{Blumen} and others, that the Green function $P(x,t)$
is well described by the fractional
diffusion approximation. 

First we consider a long wave length approximation,
and show the relation between this approximation and the fractional
diffusion framework. We rewrite Eq.
(\ref{eqE02}) in the form
\begin{equation}
u P(k,u)  - u \tilde{\psi}(k,u) P(k,u)=1 - \psi(u).
\label{egeg1}
\end{equation}
Due to Tauberian theorems, the behavior of $P(x,t)$
for $t \to \infty$ is controlled by the behavior of 
$P(k,u)$ at $u \to 0$, hence in the small $u$ limit
(and fixed $k$) one finds
$$ P(k,u)\left\{1 - \left[\tilde{\psi}(k,u)|_{u=0} - u \int_0^\infty t \cos\left( k t\right) \psi(t) dt  + \cdots \right]\right\}= $$
\begin{equation}
 \tau - Bu^{\beta -1} + \cdots .
\label{egeg2}
\end{equation}
We now consider the $k \to 0$ limit using
$ \tilde{\psi}(k,u)|_{u=0} = 1 + B |k|^{\beta} \cos\left( \pi \beta/2\right)+ \cdots$ leading to
\begin{equation}
 P(k,u)  \approx {1\over u  +B |k|^{\beta} |\cos\left( \pi \beta/2 \right)|/ \tau}.
\label{egeg3}
\end{equation}
Eq. (\ref{egeg3}) is rewritten in terms of a fractional
diffusion equation using convenient units
\begin{equation}
{\partial P_{fr}(x,t) \over \partial t} - {1 \over 2} {\partial^{\beta} \over \partial |x|^{\beta} } P_{fr}(x,t)=  \delta(x) \delta(t).
\label{egeg4}
\end{equation}
Eq. (\ref{egeg4}) describes L\'evy flights, whose solution is
a symmetric  L\'evy stable PDF given in Appendix A, Eq. 
(\ref{eqAp03}).
The derivation of Eq. (\ref{egeg4}) based on the long wave length
approximation is not rigorous; however
 Zumofen et al \cite{Blumen}
used a numerical inverse Fourier--Laplace
technique
to show that the propagator is well described by a L\'evy stable
PDF. 
This result was verified by several authors, Araujo et al \cite{Araujo}
used a numerically exact enumeration technique \cite{remark}
and Mantegna 
\cite{Mantegna} and  Weron and Weron \cite{WW}
used a Monte Carlo approach. As briefly mentioned in the introduction
Kotulsky \cite{Marcin1}
used a rigorous limit theorem  approach
to reach the same conclusion.
Thus in contradiction to the claim made in \cite{Vulp}, the
fractional equation yields a meaningful approximation to the
underlying strong anomalous diffusion process under investigation.

 We are still left with a puzzle: the fractional equation
predicts a non analytical behavior of $P(k,u)$, namely the
divergence of the even moments of the random walk,
while we know  that these moments, for any finite
time are finite. Let us therefore investigate
the moments in greater detail,
using Eq.
(\ref{eqE07z})
$ M_0(u)= 1/u$
and for $j \ge 1$,
\begin{equation}
M_{2 j} (u) \sim {B \over \tau }  \Theta_{2 j}(\beta) u^{\beta -2- 2 j},
\label{eqEn03}
\end{equation}
where $\Theta_{2 j}(\beta) \equiv \Pi_{l=0}^{2 j - 1} | l - \beta |$.
Inverting to the time domain we find the mentioned strong type of
anomalous diffusion
$M_{2 j} (t) \propto t^{2 j  + 1 - \beta}$ for $j=1,2 \cdots$
while $M_0=1$.
We now investigate the behavior of
the asymptotic moment generating function,
 based on the
method in Sec.
\ref{SecG}.
According to Eq. (\ref{eqMG})
\begin{equation}
P_{AMG}(k,u) = {1 \over u} + {B \over \tau} u^{\beta -2} \sum_{j=1}^{\infty} \left( - {k^2 \over u^2} \right)^j {\Theta_{2j}\left(\beta\right) \over \left( 2 j \right)!},
\label{eqEn04}
\end{equation}
using the identity (obtained using Mathematica)
$$  g(x) \equiv $$
\begin{equation}
\sum_{j=1}^{\infty} \left( - x
  \right)^j {\Theta_{2j}\left(\beta\right) \over \left( 2 j \right)!} =
-1 + (1 + x)^{\beta /2} \cos\left[\beta \arctan\left(\sqrt{x} \right) \right],
\label{eqEn05}
\end{equation}
we find 
\begin{equation}
P_{AMG}(k,u) = {1 \over u} + {B \over \tau } u^{\beta -2} g({k^2 \over u^2} ).
\label{eqEn06}
\end{equation}
We note that unlike the sub diffusive case,
$P_{AMG}(k,u)$ Eq. (\ref{eqEn06}) is not the Fourier Laplace transform
of the asymptotic $P(x,t)$ since $\lim_{k \to \infty}P_{AMG}(k,u)= \infty$, 
when $u$ is fixed.

To conclude,
according to long-wave length approximation
and previous work,
the jump model
propagator $P(x,t)$
is described by the fractional diffusion equation with
($\alpha=1, \mu=\beta$). This approximation does not
describe the behavior of the moments  of the underlying random walk
including the second.
These  
are described by the moment generating function Eq.
(\ref{eqEn06});
thus, two functions yield the details
on the long time behavior of the underlying random walk. 
And strong anomalous diffusion does not necessarily imply
the breakdown of the fractional approximation, though one should
take care in the interpretation of the results obtained by it.

 A similar situation occurs in the field of 
inhomogeneous line broadening \cite{Stoneham}, where the line is 
well approximated by L\'evy stable densities \cite{Barkai}
(due to long range interactions
between defects and chromophores).
However  (due to cutoffs)  even moments of
the line exist \cite{Stoneham}.
As discussed by Stoneham \cite{Stoneham}
these moments are sensitive
to behavior of the line in its
wings, and hence in the usual experimental
situation (in the field of line broadening) are not considered
relevant. 

\subsection{Ballistic Diffusion}

We now {\em  briefly} consider
the jump model with $\psi(u) = 1 - u^{\beta} \cdots $, with
$0< \beta < 1$. For this case the model exhibits
ballistic diffusion $\langle x^2(t) \rangle \propto t^2$ \cite{Blumen}.
Without going into details,
we find the asymptotic moment generating function
\begin{equation}
P_{AMG}(k,u) = { 2 u^{\beta -1} \over \left( u + i k \right)^{\beta} + 
\left( u - i k \right)^{\beta} }.
\label{eqE07}
\end{equation}
Zumofen et al \cite{Blumen} 
used an expansion in the small parameter $u\pm i k$ and
numerical simulation for $\beta=1/2$, showing that
Eq. (\ref{eqE07}) describes well long time behavior of $P(x,t)$
(at least for $\beta=1/2$).
Here we showed that the approximation in \cite{Blumen},
yields the behavior
of the moments to infinite order.
Eq. (\ref{eqE07}) is not related to a known fractional
diffusion equation. 
Note that the fractional diffusion equation in the ballistic limit
$\mu=2,\alpha=2$ yields the wave equation.

\section{L\'evy Flights with Long Rests}
\label{SecREST}

 We now consider decoupled CTRW, where
 $\tilde{\psi}(\delta x, t) = \psi(t) f(\delta x )$, assuming the moments of
$f(\delta x)$ diverge. Clearly, now the general
 approach presented in Sec. 
\ref{SecG} breaks down  and a generalization is now considered.
We assume
\begin{equation}
f(k) = 1 - {|k|^{\mu} \over 2} + { a_{2 \mu} |k|^{2 \mu} \over 4!}
-{ a_{3 \mu} |k|^{3 \mu} \over 4!} + \cdots
\label{eqL01}
\end{equation}
and $0<\mu<2$. The coefficients $a_{j \mu}$ are called the
amplitudes of the PDF $f(\delta x)$. An example being
the symmetric stable densities $f(k)=l_{\mu, 0}(k/\sqrt{2})$ where
$l_{\mu, 0}( k) \equiv \exp( - |k|^{\mu})$. We assume as before that
\begin{equation}
\psi(u)=1 -  u^{\alpha} + c_1 u^{2 \alpha} \cdots
\ \ \   0<\alpha<1,  
\label{EqL02}
\end{equation}
for $u \to 0$, 
then using Eq. 
(\ref{Eq02S})
$$ P(k,u) = {1 \over u} \left\{ 1 - \Omega(u) {|k|^\mu \over 2} + 
 \left[6 \Omega^2(u) + a_{2 \mu} \Omega(u)\right] {|k|^{2 \mu} \over 4!} - 
\right. $$
\begin{equation}
\left. \left[  90 \Omega^3(u) + 30 a_{2 \mu} \Omega^2(u) + a_{3 \mu} \Omega(u) \right] {|k|^{3 \mu} \over 6!} \cdots  \right\}.
\label{EqL03}
\end{equation}
This equation has a structure similar to Eq. 
(\ref{Eq03S}), we see that the amplitudes $a_{j \mu}$
are natural generalizations of the moments $m_{2 j}$.
We define the amplitudes $A_{j \mu}(t)$ according to
\begin{equation}
P(k,u) = \sum_{j=0}^{\infty} (-1)^j {A_{j \mu}(u) |k|^{\mu j} \over (2 j)!},
\label{EqL04}
\end{equation}
it follows from the discussion in Sec. 
\ref{SecS} that in the limit  $u \to 0$,
$A_{j \mu}(u)\sim ( 2 j)!/(2^{j} u^{j \alpha +1})$.
We define the {\em  amplitude} generating function, in the spirit
of Eq. (\ref{eqMG}), according to 
\begin{equation}
P_{AG}(k,u) = \sum_{j=0}^{\infty} (-1)^j { |k|^{\mu j}\over 2^j u^{j \alpha +1}},
\label{eqAG}
\end{equation}
and it is easy to show that
\begin{equation}
P_{AG}(k,u) = {u^{\alpha -1} \over u^{\alpha} +|k|^{\mu}/2}.
\label{EqL05}
\end{equation}
The right hand side of Eq. (\ref{EqL05}) is the Fourier-Laplace transform
of the fractional diffusion equation
(\ref{eqIn00}).
Hence the fractional 
equation describes the amplitudes of the random 
walk in the limit $t\to \infty$. However this does not necessarily
imply that for all $x$ the corresponding $P(x,t)$ describes
the CTRW in the limit $t \to \infty$. As shown already 
in subsection
\ref{subsub}, for $\mu \le 1$ and
on
the origin {\em where $P(x,t)$ attains its maximum},  the
CTRW solution and the fractional diffusion equation
solution are  
different. This limitation of the fractional equation is related
to the fact that it is  based on a small
$k$ expansion which implies large $x$ behavior. 
Note that Eq. (\ref{EqL05}) was recently suggested by Kutner \cite{Kutner}
to describe a Weierstrass flight.

\section{Discussion}
\label{SecSum}

 The fractional diffusion equation
yields the asymptotic behavior
of coupled
and decoupled CTRWs. 
However the fractional approach has its
limitations  if compared with ordinary diffusion
approximation.
Careful analysis of the underlying random walk is needed for
a better understanding of the domain of validity of the fractional equation.
In particular the fractional approximation (for decoupled
CTRWs) breaks down at the origin $x=0$
(for $\{ 0<\alpha < 1\} \cap
\{ 0< \mu \le 1\}$). More generally
 the convergence of the
CTRW solution, at $x=0$, toward the fractional approximation may become
extremely slow.
This behavior is different from ordinary random walks
where (usually) (i) deviation from  Gaussian behavior is found
in the tails $|x| \to \infty$ (i.e., for long though finite times)
and (ii) convergence toward the fractional approximation
is typically fast.
 Also note that the fractional equation has a CTRW foundation only when
$\{ 0<\alpha\le 1\} \cap 
\{ 0< \mu \le 2\}$ 
while the regime
$\{ 1< \alpha \le \mu \le 2\}$ is not related 
to the CTRWs under investigation (see however work in
\cite{BS} for $ \{ 1< \alpha \le 2=\mu\}$). 

 If they exist, moments of the decoupled CTRW
converge in a long time limit to the behavior
predicted by the fractional equation.
For coupled memories describing L\'evy walks
the situation is
more complicated.
The fractional approximation describes the 
asymptotic long time behavior of the Green function
$P(x,t)$,  though it does not describe correctly 
the moments of the underlying random walk,
not even the second moment.
In this case we may characterize the random walk using both
$P(x,t)$ and
the asymptotic moment generating function.
These two function yield different types
of information, it is still to be seen if the asymptotic
moment generating function has any universal features.

 We note that a similar situation exists 
also for some simple random walks which
are approximated with the ordinary diffusion equation.
To see this consider as an example
the sum of $N \to \infty$ independent,
identically distributed random variables $\{ x_i \}$, $i=1,2\cdots N$.
As well known the sum $\sum_{i=1}^N x_i$ converges in a limit 
to a Gaussian behavior, provided the variance of $x_i$ exists.
Assume the variance  exists but higher order moments of $x_i$
 diverge.
Then clearly the central limit theorem, or identically
the ordinary diffusion approximation,
fails to predict correctly the behavior
of the high order moments of the random walk (i.e., the Gaussian
central limit theorem does not hold at the tails of the Green function).
The situation for the fractional diffusion equation
is similar to this case, in that it fails to
predict correctly the behavior of the moments of the L\'evy walk
(i.e., the L\'evy central limit theorem does not hold at the
tails of the Green function of the L\'evy walk, where $P(x,t)=0$ for $|x|>t$).

Note that our conclusions are valid only for
free boundary conditions, for other boundary conditions
we know little on domain of validity of the 
fractional diffusion equation especially when $\mu<2$
(see however work in \cite{Tran,Ding,Buld}).

{\bf Acknowledgment:} This research was supported in part
by a grant from the NSF. 
I thank  A. I. Saichev and G. Zumofen for helpful correspondence 
and R. Silbey for
comments on the manuscript.

\section{Appendix A}

 We investigate the solution of fractional diffusion equation,
with $\alpha\le 1$ and $0<\mu \le 2$.
Following \cite{saichev} we rewrite the solution in
Fourier-Laplace space
\begin{equation}
P_{fr}(k,u) = u^{\alpha - 1} \int_0^{\infty} d s e^{ - s \left( u^{\alpha} + |k|^{\mu}/2\right)}.
\label{eqAp01}
\end{equation}
Since the symmetric L\'evy stable probability density
$l_{\mu, 0}(x)$ and $\exp( - |k|^{\mu})$ are Fourier pairs
\begin{equation}
P_{fr}(x,u) = 2^{1/\mu} u^{\alpha -1} \int_0^{\infty} d s
 e^{ - s u^{\alpha}} l_{\mu, 0}
\left( { 2^{1/\mu} x \over s^{1 / \mu} } \right) s^{ - 1/\mu}. 
\label{eqAp02}
\end{equation}
For $\alpha =1$ we use the Laplace pair
$\exp( - s u)$ and $\delta ( t - s)$ and find
as expected 
\begin{equation}
P_{fr}(x,t) =\left( {  2^{1/\mu}  \over t } \right)^{ 1 \over \mu} l_{\mu, 0} \left( { 2^{1/\mu} x \over t^{1 /\mu}} \right),
\label{eqAp03}
\end{equation}
and when $\mu=2$ the solution is Gaussian.

We now consider $\alpha<1$ and investigate the behavior
on the origin. Using Eq. 
(\ref{eqAp02})
\begin{equation}
P_{fr}(x,u)|_{x=0} = 2^{1/\mu} u^{\alpha -1} \int_0^{\infty} d s  s^{ - 1/\mu}e^{ - s u^{\alpha}} l_{\mu, 0}
\left( x \right)|_{x=0}
\label{eqAp04}
\end{equation}
where $l_{\mu, 0}(x)|_{x=0}= \Gamma(1 /\mu)/(\mu \pi)$, we find
\begin{equation}
P_{fr}(x,u)|_{x=0} =
\left\{
\begin{array}{c c}
 2^{1/\mu} u^{\alpha/\mu -1} 
\Gamma\left(1 - 1/\mu \right) 
l_{\mu, 0}
\left(x \right)|_{x=0} 
& \mu> 1 \\
\ & \ \\
\infty  &  \mu\le 1 
\end{array}
\right.
\label{eqAp05}
\end{equation}
which when inverted yields Eq. (\ref{EqL06}).
For $\mu<1$, $P_{fr}(x,u)|_{x=0}$ is infinite due to the zero lower bound in
the integral Eq. 
(\ref{eqAp04}).

 An integral solution of the fractional diffusion equation is found
using the Laplace pair $u^{\alpha -1}\exp( - s u^{\alpha})$
and $(1/\alpha) (t/s^{1 + a/\alpha}) l_{\alpha, 0}(t/s^{1/\alpha})$ 
where $l_{\alpha, 1}(t)$ is the one sided stable density
whose Laplace transform is $\exp( - u^{\alpha})$. Hence
$$ P_{fr}(x,t) =  $$
\begin{equation}
{ 2^{1/\mu}  t  \over \alpha} \int_0^{\infty} d s s^{ - ( 1 + 1/\alpha + 1/\mu)} l_{\alpha, 1}\left( {t\over s^{1/\alpha}} \right) 
l_{\mu, 0} \left( { 2^{1 /\mu} x \over s^{1 /\mu} } \right),
\label{eqAp06}
\end{equation}
this equation being valid for $0<\alpha \le 1$ and $0< \mu \le 2$.

\section{Appendix B}

 We investigate the CTRW solution
$$ P(x,u) = {1 - e^{-u^{\alpha} }\over u} \delta(x)  + $$
\begin{equation}
\left( {1 - e^{-u^{\alpha}} \over u} \right) 
\sum_{s=1}^{\infty} e^{s u^{\alpha}} 
\left({2 \over s}\right)^{1/\mu}
l_{\mu, 0}\left( { 2^{1/\mu} x \over s^{1/\mu} } \right).
\label{eqApB01}
\end{equation}
The first term 
 can be easily handled yielding $L_{\alpha,1} (t) \delta(x)$ decaying
for long times like $\delta(x)t^{-\alpha}$. This singular
term
is neglected in the fractional diffusion approximation. 
Omitting this term we find for $x=0$,
$$ P(x,u)|_{x=0} =$$
$$ \left( {1 - e^{-u^{\alpha}} \over u} \right) 
\sum_{s=1}^{\infty} e^{-s u^{\alpha}} 
\left({2 \over s}\right)^{1/\mu} 
l_{\mu, 0}\left(x\right)|_{x=0} =$$
\begin{equation}
2^{1/\mu}
\left( {1 - e^{-u^{\alpha}} \over u} \right) 
\mbox{PolyLog}\left[ {1 \over \mu}, e^{ -u^{\alpha}} \right]
l_{\mu, 0}\left(x \right)|_{x=0}
\label{eqApB03}
\end{equation}
where $\mbox{PolyLog}\left[n,z\right]$ is the $n$ th polylogarithm
function of $z$. For $\mu<1$ and  $u \to 0$ we  find
\begin{equation}
P(x,u)|_{x=0} \sim 
2^{1/\mu}
u^{\alpha -1} 
\zeta\left({ 1 \over \mu} \right)
l_{\mu, 0}(x)|_{x=0},
\label{eqApB04}
\end{equation}
where $\zeta(z)$ is Riemann's zeta function. 
For $\mu=1$ we use
$\sum \exp( - s u^{\alpha})/s=-\ln[1 -\exp(-s u^{\alpha})]$
to find
\begin{equation}
P(x,u)|_{x=0}\sim - 2 u^{\alpha - 1} \ln\left( u^{\alpha}\right)
  l_{1,0}(x)|_{x=0}.
\label{eqApB05}
\end{equation}
For $1<\mu<2$ we use the Euler--Mclaurin summation formula
$$ \sum_{s=1}^\infty  e^{ - s u^{\alpha}} s^{ - 1/\mu} =
e^{- u^{\alpha}} \left[ 1 + 
\sum_{k=1}^\infty e^{ - k u^{\alpha}} (k + 1)^{-1/\mu}
\right] = $$
\begin{equation}
e^{ - u^{\alpha}} \left[ 1 + \int_0^{\infty} dk  {e^{ - k u^{\alpha} } \over \left( 1 + k \right)^{1/\mu} } - {1 \over 2} + {1 \over 12} {d \over dk} {e^{- ku^{\alpha}} \over \left( 1 + k \right)^{1 /\mu } } |_{0}^{\infty} + \cdots \right]
\label{eqApB06}
\end{equation}
where $\cdots$ are the higher order terms in the 
Euler--Mclaurin formula. 
The integral in Eq. (\ref{eqApB06}) in the limit $u \to 0$ yields
\begin{equation}
\int_0^{\infty} dk  {e^{ - k u^{\alpha} } \over \left(
 1 + k \right)^{1/\mu} } \sim u^{\alpha / \mu - \alpha} \Gamma\left(1 - {1 \over \mu} \right),
\label{eqApB07}
\end{equation}
provided that $\mu>1$. This term is much larger than the other terms in
Eq. 
(\ref{eqApB06}); hence, we find
\begin{equation}
P(x,u)|_{x=0} \sim 
2^{1 / \mu} u^{ \alpha/\mu -1} \Gamma\left(1 - {1 \over \mu} \right)
l_{\mu, 0}(x)|_{x=0}. 
\label{eqApB08}
\end{equation}
Inverting Eqs. (\ref{eqApB04},\ref{eqApB05},\ref{eqApB08}) we find Eq. 
(\ref{eqEx06}).


\begin{thebibliography}{99}
\bibitem{remark2} 
Closed form expressions for
L\'evy stable PDFs, were obtained by Schneider \cite{Sch1},
 in terms of Fox H functions
(i.e., for $0<\mu<2$ and $0<\alpha<1$).
However, only for special cases one finds a simple (i.e., tabulated
and accessible) closed form
expression for L\'evy stable probability densities.
Series expansions,
and asymptotic behaviors of stable laws are summarized in \cite{Sch1}.
Note that \cite{Sch1} also points out relevant errors in the literature.
A brief summary of properties of one sided stable
laws can also be found in an Appendix in \cite{Tran}.
\end{thebibliography}
\end{document}